# Spectroscopic investigation of native defects induced electron-phonon coupling in GaN nanowires

Santanu Parida,[1] Avinash Patsha,[1] Santanu Bera[2] and Sandip Dhara[1]

[1]*Nanomaterials Characterisation and Sensors Section, Surface and Nanoscience Division, Indira Gandhi Centre for Atomic Research, Homi Bhabha National Institute, Kalpakkam-603102, India*

[2]*Water and Steam Chemistry Division, BARC Facility, Kalpakkam-603102 & Homi Bhabha National Institute, Mumbai-400094, India.*

E- mail: avinash.phy@gmail.com and dhara@igcar.gov.in

**Abstract**

The integration of advanced optoelectronic properties in nanoscale devices of group III nitride can be realized by understanding the coupling of charge carriers with optical excitations in these nanostructures. The native defect induced electron-phonon coupling in GaN nanowires are reported using various spectroscopic studies. The GaN nanowires having different native defects are grown in atmospheric pressure chemical vapor deposition technique. X-ray photoelectron spectroscopic analysis revealed the variation of Ga/N ratios in nanowires having possible native defects, with respect to their growth parameters. The analysis of characteristics features of electron-phonon coupling in Raman spectra show the variations in carrier density and mobility with respect to the native defects in unintentionally doped GaN nanowires. The radiative recombination of donor acceptor pair transitions and the corresponding LO phonon replicas observed in photoluminescence studies further emphasize the role of native defects in electron-phonon coupling.

Keywords: electron-phonon coupling, native defects, Raman spectroscopy, XPS, photoluminescence

# 1. Introduction

Group III nitrides attract remarkable attention because of their tunable electronic and optical properties by appropriate alloy formation. Due to the extensive reduction of extended defects in 1D nanostructures, III-nitride based nanowires (NWs) have emerged as ideal candidates for building blocks of nanoscale electronic and optoelectronic devices such as LEDs, laser diodes, photodetectors ranging from infrared to ultraviolet regime [1-3]. The III-nitride NW heterojunctions also find application for high electron mobility transistor and high efficiency solar cells [4,5]. Moreover, recently the group III nitride NWs based on GaN gained much attention because of its capability to promote hydrogen generation by photo-catalytic water splitting [6]. The performance of such highly efficient nanoscale electronic and optoelectronic semiconductor devices also depends on the dopants and defects, which control the free carrier density as well as carrier mobility. Among the commonly known native point defects such as N vacancy ($V_N$), O antisite in place of N ($O_N$) and Ga vacancy ($V_{Ga}$) in unintentionally doped GaN, $O_N$ is ubiquitous inevitable impurity in group III nitride. The concentration of unintentional O impurity in GaN films is found to be order of $10^{17}$–$10^{21}$ /cm$^3$, which results in the net carrier concentration of $10^{16}$ – $10^{20}$ /cm$^3$ [7,8]. The incorporation of native defects in GaN NWs can be due to growth process leading to the efficiency in the supply of precursor materials (Ga/N ratio), and crystallographic orientation. Several studies were carried out to realize the effect of supply of Ga/N ratio and O impurity incorporation on stoichiometry, size, surface morphology, growth rate, optical and electrical properties of GaN NWs [9-13]. Unlike in the case of thin films, it is difficult to perform measurements like secondary ion mass spectroscopy, Hall probe on single nanowires to quantify the concentration of native defects, resultant net carriers, and their mobility due to the nanowire geometry.

Therefore, it is necessary to explore and understand the utilization of alternative spectroscopic techniques for the study of native defects in nanowires.

The optical and electrical properties of semiconducting materials are well known to be influenced by the basic interactions of exciton or charge carriers with lattice vibrations. The reduction in carrier mobility, broadened near edge optical transitions, and cooling of hot carriers are among the direct consequences of electron-phonon interactions [14]. Different kinds of interactions are possible with different lattice vibrations in GaN. The nonpolar transverse optical (TO) phonons, interact with electrons by means of deformation–potential (DP), while the polar longitudinal optical (LO) phonons interact with electrons by means Coulomb interactions known as Fröhlich interaction (FI) [14,15]. The electron-phonon interactions in GaN films are well studied by means of Raman and photoluminescence (PL) spectroscopy analysis. The plasmon coupled LO phonon modes are studied by analyzing Raman spectra of GaN, revealing the strong electron-phonon coupling in the system with different carrier density [16-19]. Most of the studies have been carried out by varying the extrinsic doping concentration in GaN. Nevertheless, there are hardly any reports on the effect of native defects ($V_N$, $O_N$, $V_{Ga}$) in unintentionally doped GaN and resultant free carriers on electron-phonon coupling in GaN nanowires using optical spectroscopy. The same is also missing for the study of the strength of interaction potentials in the analysis of carrier concentration and mobility. Moreover, the PL properties of GaN are also reported to be influenced in the presence of electron-phonon coupling, which may be further modified by the formation of point defects [13,20]. In addition, the morphology and size of NWs can also modify the electron-phonon interactions depending on the surface architecture [21,22]. Thus, the understanding of electron-phonon interactions is critical for manipulating the optical and electrical properties of GaN NWs.

Therefore, the present study aims at shedding light on the various electron-phonon coupling characteristics in unintentionally doped GaN NWs having different native defect density. In order to manipulate the defect induced free charge carriers, GaN NWs were grown using atmospheric pressure chemical vapor deposition (APCVD) technique under variable precursor flow rates. Detailed Raman and PL spectroscopic studies were performed to explore the electron-phonon interactions induced by native defects whose existences are corroborated by the X-ray photoelectron spectroscopic studies.

**2. Experimental details**

*2.1. Synthesis of GaN nanowires*

GaN NWs were synthesized by APCVD technique using vapor-liquid-solid (VLS) process. Au catalyst was deposited on crystalline (*c*-)Si(100) substrates using thermal evaporation technique (12A4D, HINDHIVAC, India). These substrates were annealed for making the Au nanoparticles (NPs) at a temperature of 900 ℃ for 10 min in an inert atmosphere. Ga metal (99.999%, Alfa Aesar) as precursor, $NH_3$ (5 N pure) as reactant and Ar (5 N pure) as carrier gases were used for the growth process. The Si(100) substrate with Au NPs was kept upstream of a Ga droplet (~200 mg) in a high pure alumina boat (99.95%), placed inside a quartz tube. The temperature of the quartz tube was raised to a growth temperature of 900 ℃ with a ramp rate of 15 ℃ per min. Growth was carried out for 180 min by purging $NH_3$, with different flow rates and keeping a constant amount of Ga (200 mg) in the reaction chamber. Sample S1 was grown under N rich condition by flowing 50 sccm of $NH_3$ without any carrier gas. On the other hand, sample S2 and S3 were grown under N reduced environment by reducing the flow rate of $NH_3$ to10 sccm and diluted with the carrier Ar gas of 10 and 20 sccm, respectively.

*2.2. Characterization*

Morphological study was carried out by using a field emission scanning electron microscope (FESEM; AURIGA, Zeiss). Compositional analysis was performed by utilizing X-ray photoelectron spectrometer (XPS; VG, ESCALAB MK200X) using an X-ray source of Al-Kα (1486.6 eV), and the binding energy (BE) values were measured with respect to the C1s reference peak. The X-ray beam diameter used for the XPS measurements was around 3 mm and the collection area (with the largest slit) for the analysis was approximately 2×3 mm$^2$. The spectra were processed by applying Shirley type background and curve fitted with mixture of Gaussian–Lorentzian line shapes. In order to understand the phase formation, vibrational spectra were collected using a micro-Raman spectrometer (inVia, Renishaw; UK) in the backscattering geometry with a Ar$^+$ laser excitation of 514.5 nm. A grating having 1800 groves·mm$^{-1}$ as a monochromatizer and thermoelectrically cooled charge coupled device (CCD) as a detector were used for recording the spectra. The spectra were acquired using a 50× objective with numerical aperture of 0.75. The PL spectroscopy was carried out by exciting the samples with a He-Cd laser excitation of 325 nm. The scattered photons were collected by a 15X UV objective with numerical aperture of 0.32 and dispersed through a grating of 2400 groves·mm$^{-1}$ to the CCD. The spectra were recorded at both room temperature and 80K using the liquid N$_2$ cooled cryostat (Linkam THMS 600).

**3. Results and discussions**

*3.1. Morphological characteristics*

The typical FESEM images of as-grown GaN NWs from sample S1-S3 are shown (figure 1a-c). The Au NPs at the tip show that the NWs are grown in the VLS process. The FESEM micrograph of sample grown under N rich condition (S1) shows very rough and non-uniform surface morphology for the as-grown NWs having a size distribution of ~80-120 nm with a less growth rate of ~ 0.5 µm/hr (figure 1a). On the other hand, the NWs grown under N reduced condition (S2 and S3) show a quite smooth and uniform surface morphology with a growth rate ~2 µm/hr and size distribution of ~60-120 and ~60-150 nm for S2 and S3 respectively (figure 1b and c). The reduction in growth rate for sample S1 can be attributed to the self-nitridation of Ga source in case of N rich condition where growth rate is impeded with the formation of a thin nitride layer on the top surface of the Ga source it-self and the observation is consistent with the earlier report [10].

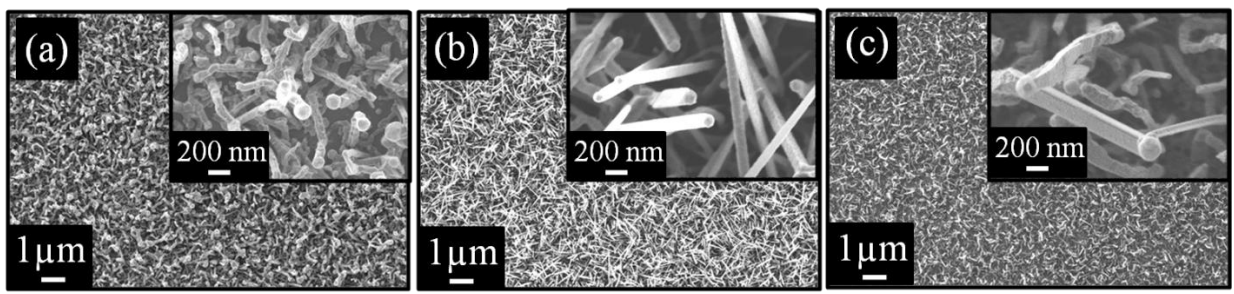

Figure 1. Typical FESEM image of GaN NWs of sample (a) S1 (b) S2, and (c) S3. Insets showing the high magnification images of as-grown NWs.

*3.2. XPS investigations for compositional analysis*

The XPS study was carried out to estimate the Ga/N ratio in NWs for all the samples. Typical XPS spectra for the samples are shown in the figure 2 for different elements with their characteristic electronic transitions. The peak centered ~19.4 eV corresponds to binding energy of electron in 3d electronic state of Ga. N 1s level transition was observed at ~397 eV from the de-convolution of a broad peak, which is considered as a combination of N 1s and Ga LMM Auger transitions. Along with Ga and N, a characteristic peak of O 1s was also observed at ~531 eV [23,24]. The presence of O peak in XPS shows that the unintentional incorporation of O in GaN may have occurred during the APCVD growth process, as the base vacuum of the chamber and purging of precursor gases are inadequate to reduce the presence of O. Other sources like substrate and quartz tube can also contribute to the O contamination. However, the presence $SiO_x$ layer on Si which was used as a substrate to grow the GaN NWs might have also contributed to the observed atomic percentage of O in all the samples, as the NWs have not fully covered the Si substrate. However, the relative variation of O atomic percentages (at.%) from sample S1 to S3 represents its contribution from the GaN lattice. The at.% of Ga, N and O are calculated from the area under the curves and are tabulated in table 1.

**Table 1.** Atomic percentage of the elements calculated from XPS spectra

| Sample | Ga (at. %) | N (at. %) | O (at. %) | Ga/N ratio |
|---|---|---|---|---|
| S1 | 29.9 | 41.8 | 28.3 | 0.715 |
| S2 | 27.9 | 40.1 | 32.0 | 0.696 |
| S3 | 26.7 | 39.9 | 33.4 | 0.670 |

XPS analysis shows that the decrease in both Ga and N at.% and increase in O at.% from sample S1 to S3. Surprisingly, less atomic percentage of Ga was observed in the sample S3, grown under N reduced, (i.e. Ga rich) condition. However, the observed increase in O at.% from sample S1 to S3 could be due to the reduction in the N concentration during their growth. It may be emphasized that since S3 is grown under less concentration of $NH_3$, the amount of O involved in the growth process is expected to be high by forming point defects $O_N$ and $2O_N$ defect complex in GaN [12]. Furthermore, the excess $O_N$ defects enhance the $V_{Ga}$ density by forming $V_{Ga}$-$3O_N$ defect complex [12], and hence a reduction in the Ga at.% is observed in the XPS analysis. Moreover, we calculated the Ga to N ratio (Ga/N) in all the samples and the ratio for sample S1 (~0.715) is found to be close to the stoichiometric value of unity, proposing a good crystalline quality and hence, the optical property in the sample.

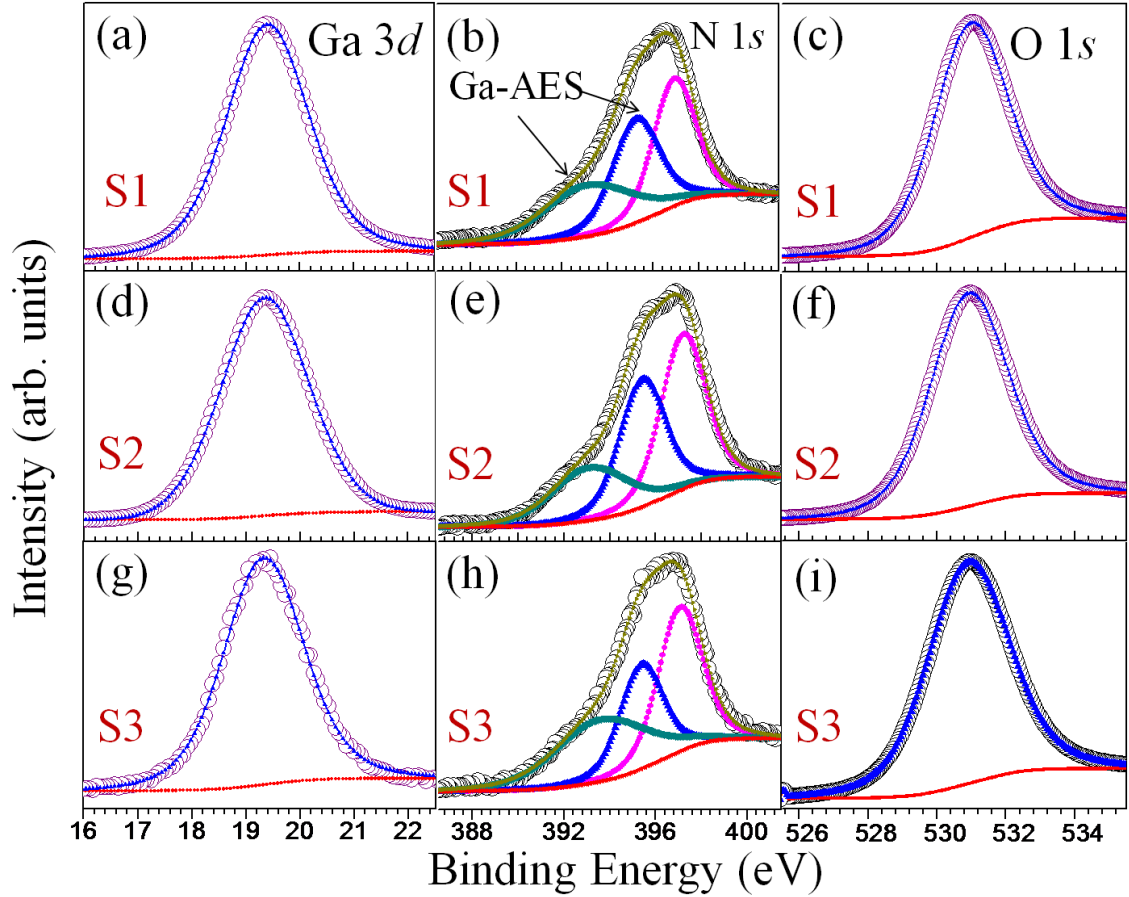

Figure 2. Typical XPS spectra of different elements and its characteristic electronic transitions for samples (a)-(c) S1, (d)-(f) S2 and (g)-(i) S3.

*3.3. Characteristics of electron-phonon coupling in Raman spectra*

The typical Raman spectra are shown (figure 3) for the NW samples grown under different N concentration. The first-order Raman scattering is caused by phonons with wave vector $k \sim 0$ because of the momentum conservation rule in the inelastic light scattering process. In case of a wurtzite hexagonal structure, group theory predicts eight sets of phonon normal modes at the Γ-point, namely, $2A_1 + 2E_1 + 2B_1 + 2E_2$. Among them, one set of $A_1$ and $E_1$ symmetry modes are acoustic, while the remaining six modes ($A_1 + E_1 + 2B_1 + 2E_2$) are optical. Each one of $A_1$ and $E_1$ optical modes split into LO and transverse optical (TO) modes due to their polar nature of vibration [19,25,26].

The Raman spectra of NWs from all the samples were fitted with Lorentzian line shape and were analyzed for the corresponding phonon modes. As grown NWs of sample S1 (figure 3a) show the peaks centered at ~567, ~537, and 721 cm$^{-1}$ corresponding to Raman active symmetry allowed $E_2$(high) phonon mode, transverse and longitudinal optical (TO and LO) phonon modes of $A_1$ symmetry ($A_1$(TO)) and $A_1$(LO)), respectively, confirming presence of wurzite GaN phase [19,25,26]. Along with the Raman active modes two broad peaks centered ~628 and ~680 cm$^{-1}$ are assigned to the surface optical (SO) phonon modes of GaN corresponding to $A_1$ and $E_1$ symmetries, respectively [21]. An intense peak centered at ~521 cm$^{-1}$ corresponds to the Si substrate on which GaN NWs were grown (figure 3a). The Raman spectra of GaN NWs of sample S2 (figure 3b) also showed similar characteristics as that of S1, along with a new peak centered at ~420 cm$^{-1}$ which corresponds to the zone boundary (ZB) phonon mode of GaN arising due to the finite size of the crystal [21]. In addition, the $A_1$(LO) symmetry phonon mode is blue shifted by ~3 cm$^{-1}$. Whereas for the NWs of sample S3 (figure 3c), along with an intense peak of $E_2$(high) phonon mode, a peak corresponds to the TO mode of $E_1$ symmetry ($E_1$(TO)) at ~551 cm$^{-1}$ is also observed. The $A_1$(LO) mode is further blue shifted by 6 cm$^{-1}$.

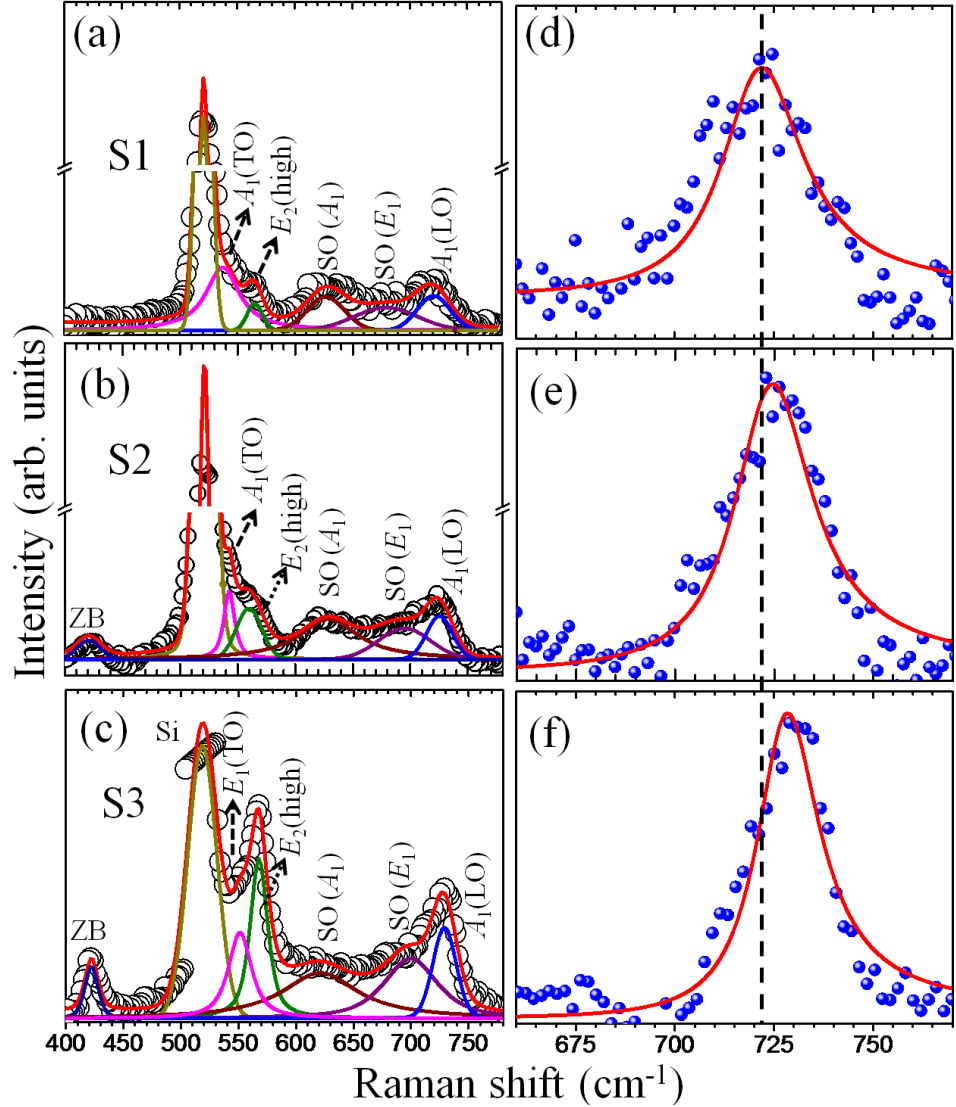

Figure 3. Typical Raman spectra of GaN NWs using 514.5 nm excitation from samples (a) S1, (b) S2 and (c) S3 with Lorentzian line shape fitted. (d)-(f) Corresponding $A_1$(LO) mode fitted with line shape analysis. Vertical dashed line is a guide to eye for the observed blue shift of $A_1$(LO) mode.

*3.3.1. Fröhlich and deformation-potential interactions involved electron-phonon coupling*

As discussed earlier, the electron–phonon coupling can occur either by FI and/or by DP interactions. The non-polar phonons, such as $E_2$, TO modes are mostly involved in interacting with DP, whereas polar LO phonons are involved in both FI and DP interactions [14]. In the ionic crystals like GaN, the FI is dominant depending on the excitation energy. The relative strength of both FI and DP interactions by different phonon modes can be revealed by considering the ratio of Raman intensities of corresponding phonon modes [27]. Since the matrix element $\{\langle k|H_{ep}|j\rangle\}$ of electron-phonon interaction ($H_{ep}$) in Raman intensity is different for different phonon modes, the ratio of Raman intensities of particular phonon modes is equivalent to the ratio of magnitudes of the corresponding matrix elements as given by eqn. (1), and hence the relative strength of the interactions [27].

$$(I_{E2,TO} / I_{LO}) \approx \{ |\langle k|H_{ep}(E_{2,TO})|j\rangle|^2 / |\langle k|H_{ep}(LO)|j\rangle|^2 \} \approx \{ |\langle k| DP|j\rangle|^2 / |\langle k|(DP+FI)|j\rangle|^2 \} \ldots\ldots\ldots(1)$$

Variations are often observed in relative Raman intensities of $E_2$(high), $E_1$(TO), $A_1$(TO) and $A_1$(LO) phonon modes in GaN, with respect to crystal orientation, collection geometry for scattered photons, excitation energy, and presence of dopants [16-19,28]. In the present studies, although the scattered photons from GaN NWs are collected in unpolarized condition and the NWs are aligned randomly on substrate, the relative Raman intensities of $E_2$(high) and $A_1$(LO) phonon modes are found to vary among the three NWs samples (figures 3a-c). Such variations in the Raman intensity could be due to the relative strength of the electron-phonon interactions. In order to understand the strength of these interactions, the ratio of intensities of $E_2$(high) and LO phonon modes ($I_{E2}/I_{A1-LO}$), and the ratio of intensities of TO and LO phonons ($I_{TO}/I_{LO}$) are estimated and compared among the three samples. The ratios ($I_{E2}/I_{A1-LO}$) for S1, S2, and S3 are found to be 0.3, 1.0, and 1.76, respectively. Whereas, the ratios ($I_{A1-TO}/I_{A1-LO}$) for S1 and S2 are 1.80 and 1.57, respectively. For the sample S3, the ratio ($I_{E1-TO}/I_{A1-LO}$) is estimated to be 0.94. The values suggest that DP interactions with $E_2$(high) phonons is stronger in case of sample S3 than that of the sample S1. While in case of non-polar TO phonons in III-nitrides [14], the contribution of DP interaction is significantly reduced as compared to $E_2$(high) mode leading to an overall decrease in the ratio $I_{TO}/I_{LO}$. In this context it may be noted that an approximately six-fold increase is observed for the ratio $I_{E2}/I_{A1-LO}$ indicating a significant role of DP in sample S3 (supplementary figure S1 showing excitation dependent Raman studies for the determination of resonance condition).

*3.3.2. Defect induced free carrier plasmon–phonon coupling*

The observed shift in $A_1$(LO) phonon mode of NWs from samples S1 to S3 (figures 3d-f) could be due to the interactions among LO phonons and free charge carriers in the samples, as such shift is not possible in $A_1$(LO) phonons alone if it is associated with strain, or temperature effects [19]. Further, considering the similar experimental conditions for all the samples in the present study with identical lattice mismatch due to Si substrate and a constant low power laser excitation of <1 mW, the shift cannot be attributed to the strain or temperature effect. However, for semiconductors, especially in ionic solids the polar LO phonons strongly interact with the free charge carriers such as electrons present in the system by means of Fröhlich interaction. As revealed by the XPS studies the GaN NWs are expected to possess different carrier concentration as a function of native defects present in the system [8,10]. The defect induced free charge carriers thus can strongly interact with LO phonons, resulting in the blue shift of $A_1$(LO) phonon modes from samples S1 to S3. By analyzing the line shape of the coupled plasmon-phonon modes, it has been shown that the free carrier density can be calculated, considering the DP and FI in Raman scattering cross-section [19,29].

The Raman scattering intensity ($I(\omega)$) as a function of phonon ($\omega$) and plasmon ($\omega_p$) frequency is described by

$$I(\omega) \propto A(\omega) \times \text{Im}\left[\frac{-1}{\varepsilon(\omega)}\right] \quad\quad\quad\quad\quad\quad\quad\quad (2)$$

where $A(\omega)$ is the correction factor by DP and FI in terms of Faust–Henry coefficient, TO phonon ($\omega_{TO}$) and LO phonon ($\omega_{LO}$) frequencies, plasmon damping constant ($\gamma$), and phonon damping constant ($\Gamma$), and $\varepsilon(\omega)$ being the dielectric function described as

$$\varepsilon(\omega) = \varepsilon_\infty \times \left\{1 + \frac{\omega_{LO}^2 - \omega_{TO}^2}{\omega_{TO}^2 - \omega - i\Gamma\omega} - \frac{\omega_p^2}{\omega^2 + i\gamma\omega}\right\} \quad\quad\quad\quad\quad (3)$$

The line shape of $A_1(LO)$ phonon modes of the three NWs samples (figures 3d-f) were fitted using the eqn. (2), by adjusting the fitting parameters $\omega_p$, $\gamma$, $\Gamma$, and C. From the fitted values of $\omega_p$, $\gamma$ and using the effective mass of the electron ($m^*$) in GaN, the carrier concentration $\{n_e = (\omega_p^2 \varepsilon_\infty m^*)/4\pi e^2\}$ and mobility $\{\mu_e = e/m^*\gamma\}$ of the NWs were estimated and tabulated (Table 2).

**Table 2.** Carrier concentration and mobility of the samples calculated from line shape analysis of the coupled LO phonon mode.

| Sample | Ga/N ratio | $n_e$ (cm$^{-3}$) | $\mu_e$ (cm$^2$.V$^{-1}$.s$^{-1}$) |
|---|---|---|---|
| S1 | 0.715 | $4.4\times10^{16}$ | 443 |
| S2 | 0.696 | $1.6\times10^{17}$ | 234 |
| S3 | 0.670 | $6.8\times10^{17}$ | 57 |

The estimated $n_e$ and $\mu_e$ values of the three samples show the increase in $n_e$ from $4.4\times10^{16}$ cm$^{-3}$ for sample S1 to $6.8\times10^{17}$ cm$^{-3}$ for S3, whereas $\mu_e$ is found to decrease from 443 cm$^2$.V$^{-1}$.s$^{-1}$ for S1 to 57 cm$^2$.V$^{-1}$.s$^{-1}$ for S3. Although, the NWs are not extrinsically doped with any dopants such as Si (*n*-type dopant), the increase in the values of ne further confirm the presence of native donor defects such as $V_N$ and $O_N$. In GaN, the native defects like $V_{Ga}$, $V_N$, $O_N$ and the defect complexes $2O_N$, $V_{Ga}$-$O_N$, and $V_{Ga}$-$3O_N$ are known to form acceptor, donor or double donor states, respectively [12,30]. The increase in the density of donor states over acceptors is thus responsible for increase in the free carrier concentration for the NWs of S3 as compared to that of S1. Similarly, the reduction in mobility of GaN NWs from S1 to S3 can also be attributed to the increase in the defect density in S3 as compared to that of the S1. Furthermore, the variations of $n_e$ and $\mu_e$ with respect to Ga/N ratios in NWs are analysed (figure 4) to understand the role of native defects in electron-phonon coupling with respect to growth parameters of NWs. The increase in carrier concentration with decreasing the Ga/N ratio and the decrease in mobility with decreasing Ga/N ratios were observed in the GaN NWs. The NWs having Ga/N ratio close to the stoichiometry values (0.715) are found to have lowest carrier concentration ($4.4\times10^{16}$ cm$^{-3}$) and highest mobility (443 cm$^2$.V$^{-1}$.s$^{-1}$). On the other hand, the NWs having lowest Ga/N ratio (0.670) are found possess the large carrier concentration ($6.8\times10^{17}$cm$^{-3}$) with very low mobility (57cm$^2$.V$^{-1}$.s$^{-1}$). The off-stoichiometry of Ga/N ratio in NWs enhanced the presence of native defects $V_N$, $V_{Ga}$ which contributed to the resultant free charge carrier concentration in the NWs. In addition, the increase in O impurities with off-stoichiometry further enhanced the carrier density due to formation of $O_N$ donors, resulting in the large net charge carrier concentration in NWs of sample S3. Due to enhanced scattering among large carrier density and scatting between free carriers and native defects in NWs, the charge mobility is lowest in the NWs having off-stoichiometry.

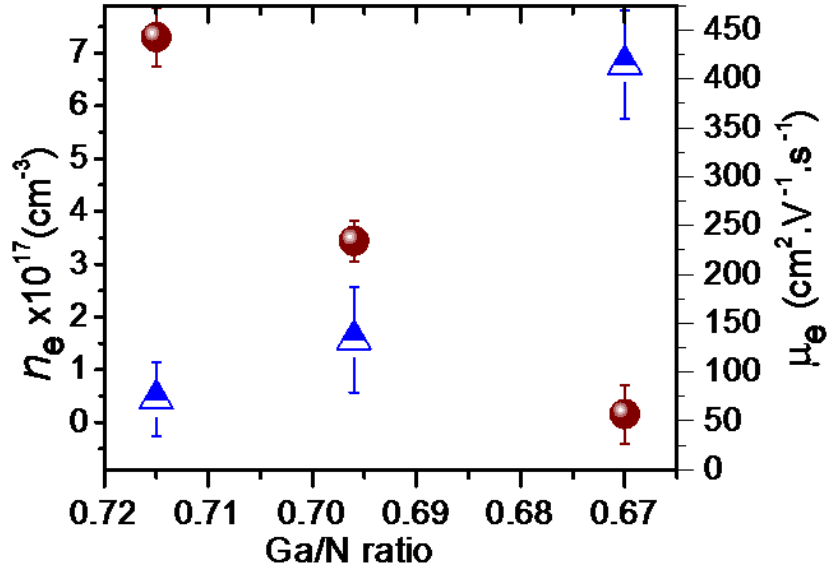

Figure 4. Variation of free carrier density and carrier mobility with respect to Ga/N ratio in GaN NWs.

The XPS results of Ga/N ratio and O impurities in GaN NWs, and the Raman spectroscopic analysis of variations in $n_e$ and $\mu_e$, strongly emphasizes the role of native defects in electron-phonon coupling which can vary with growth parameters, even without the extrinsic doping in NWs.

*3.4. Native defects involved optical transitions and corresponding electron–phonon coupling characteristics.*

The presence of native defects and their role in electron-phonon coupling during the radiative transitions in GaN NWs were investigated using PL spectroscopy. A UV laser of 325 nm wavelength was used to excite the nanowires and the emission spectra were recorded at room temperature (300 K) and at 80 K. The PL spectra of samples S1, S2, and S3 collected at 300 K (figures 5a-c), and 80 K (figures 5d-f) are shown. The luminescence from the NWs of sample S1 at 300 K, shows a broad band between 3.0 - 3.6 eV (figure 5a). A clear splitting of the broad band is observed on cooling the sample to 80 K (figure 5d). The peak around 3.51 eV corresponds to the emission due to the recombination of electron-hole pairs from a free exciton (FE). Along with the FE emission, the emission peaks around 3.47, and 3.27 eV are also observed in the PL spectra. The peak at ~3.47 eV corresponds to the emission due to radiative recombination of electron-hole pairs from excitons, bound to the sallow donor (DBE) states created by $V_N$ and $O_N$ [13,20]. The peak around 3.27 eV is attributed to recombination of the donor–acceptor pair (DAP), due to a transition of electrons from a shallow donor state ($V_N$, $O_N$) to a deep acceptor state of $V_{Ga}$. A broad peak between 2.7 - 3.0 eV, in the low temperature PL spectra (figure 5d) corresponds to the blue luminescence (BL), which is often reported in undoped GaN and is assigned to the transitions involved with defect states of shallow acceptors ($V_{Ga}$) and deep donors ($V_N$) [20].

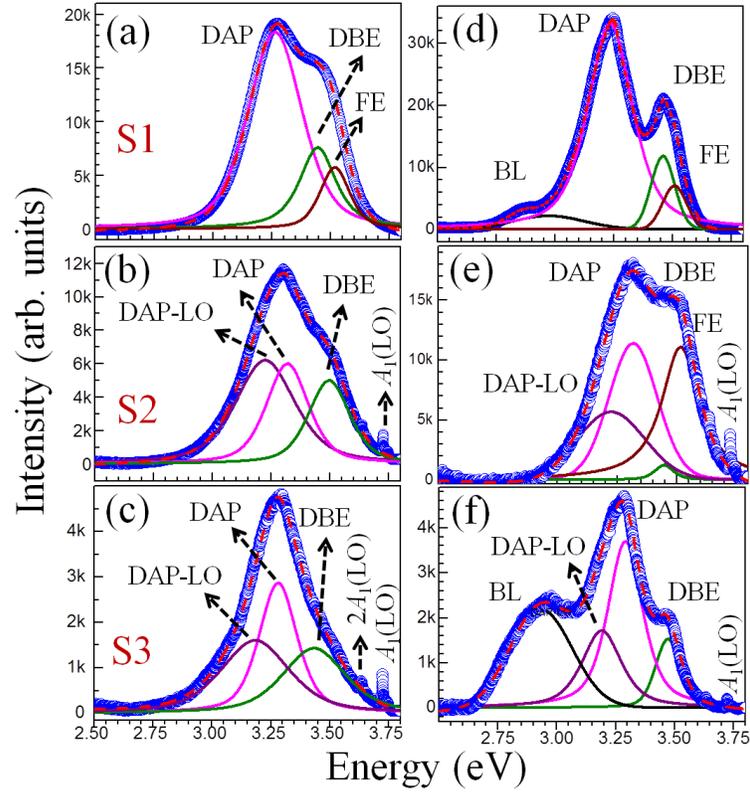

Figure 5. Typical photoluminescence spectra of GaN NWs of samples (a) S1, (b) S2 and (c) S3, collected at room temperature and their corresponding low temperature (80K) spectra (d)-(f) respectively.

The PL spectra of sample S2 show a FE emission ~3.51 eV along with DAP (~3.31) peak which is blue shifted by 40 meV as compared to that of the sample S1 at 80K (figure 5e). The new emission peak observed around 3.22 eV, which is ~ 90 meV less in energy as compared to the DAP peak, is attributed to the LO phonon replica of DAP (DAP-LO) [20]. In addition to the LO-phonon replica of DAP, a sharp resonance Raman line of $A_1$(LO) phonons in PL spectra further confirmed the electron-phonon coupling, as observed in Raman spectroscopic investigations of the samples. The sample S3 emitted the luminescence due to DBE (3.47 eV), and a DAP transition along with its LO phonon replica (DAP-LO) around 3.19 eV (figures 5c and 5f). However, the luminescence due to FE emission is not observed even at 80 K. The spectra also showed a strong BL peak around 2.9 eV at low temperature. The quenching in luminescence originating due to FE emission is observed with decrease in Ga/N ratio of NWs sample S1 to S3.

The PL study provides a clear evidence of the presence of native defects ($V_N$, $O_N$, and $V_{Ga}$) in GaN NWs having different Ga/N ratio as supported by XPS analysis. Since, the carrier concentration is expected to be higher in the samples grown under N reduced condition (S2 and S3), which favors the formation of native defects, the electron-phonon interaction is strengthened in these samples and consequently a phonon replica (DAP-LO) was observed in the PL spectra collected at both room temperature (figures 5b and 5c) and low temperature (figures 5d and 5e). In general, FE and DBE cannot be observed for semiconductors at room temperature. However, we can observe these luminescent peaks in case of GaN, as the exciton bound to the donor (arising due to N vacancy ($V_N$)) is having energy ~64(±10) meV [31], which is far above the room temperature thermal energy of 25 meV. The PL spectra collected from samples S2 and S3 also show Raman mode corresponding to $A_1$(LO) phonon along with its second order mode both at room temperature (figures 5b and 5c) and low temperature as well. However, there is no signature of LO phonon modes in the PL spectra of sample S1. The observation of optical phonon modes in PL spectra of samples S2 and S3, further confirms the strong electron–phonon coupling induced by Fröhlich interaction near resonance condition, in the presence of large number of carriers originating from native defects in these NWs.

## 4. Conclusions

GaN nanowires (NWs) having different Ga/N ratio were grown by changing the flow rate of $NH_3$ in atmospheric pressure chemical vapor deposition technique. Compositional analysis revealed the presence of different Ga/N ratios in the samples, with sample grown under N rich condition showing composition close to the stoichiometric ratio. Raman spectroscopic analysis showed different strength of interaction of phonon modes with the lattice deformation potential (DP) and Coulomb type Fröhlich interaction (FI) for the samples with different Ga/N ratio. The influence of native defect induced carriers on Raman spectra were confirmed from the observed blue shift of $A_1$(LO) mode with decreasing Ga/N ratio for the sample grown under N reduced condition. The carrier concentration and mobility of the GaN NWs were calculated from the line shape analysis of the plasmon coupled phonon modes for the samples with different Ga/N ratio. The sample with least value of Ga/N ratio showed the highest carrier concentration and the lowest carrier mobility as compared to that the sample close to the stoichiometry, caused by the presence of higher native defect density in the former. Photoluminescence study showed the phonon replica, as well as both the first and the second order optical $A_1$(LO) phonon modes, which further confirm the presence of strong electron-phonon coupling for the samples grown under N reduced conditions.

**Acknowledgements**

We thank R. Pandian of SND, IGCAR, for his help in the FESEM study. We also thank A. K. Sivadasan, Kishore K. Madapu and R. Basu of SND, IGCAR, for their useful discussions.

**Table 1.** Atomic percentage of the elements calculated from XPS spectra

| Sample | Ga (at. %) | N (at. %) | O (at. %) | Ga/N ratio |
|---|---|---|---|---|
| S1 | 29.9 | 41.8 | 28.3 | 0.715 |
| S2 | 27.9 | 40.1 | 32.0 | 0.696 |
| S3 | 26.7 | 39.9 | 33.4 | 0.670 |

**Table 2.** Carrier concentration and mobility of the samples calculated from line shape analysis of the coupled LO phonon mode.

| Sample | Ga/N ratio | $n_e$ (cm$^{-3}$) | $\mu_e$ (cm$^2$.V$^{-1}$.s$^{-1}$) |
|---|---|---|---|
| S1 | 0.715 | $4.4 \times 10^{16}$ | 443 |
| S2 | 0.696 | $1.6 \times 10^{17}$ | 234 |
| S3 | 0.670 | $6.8 \times 10^{17}$ | 57 |

**Figure Captions:**

Figure 1. Typical FESEM image of GaN NWs of sample (a) S1 (b) S2, and (c) S3. Insets showing the high magnification images of as-grown NWs.

Figure 2. Typical XPS spectra of different elements and its characteristic electronic transitions for samples (a)-(c) S1, (d)-(f) S2 and (g)-(i) S3.

Figure 3. Typical Raman spectra of GaN NWs using 514.5 nm excitation from samples (a) S1, (b) S2 and (c) S3 with Lorentzian line shape fitted. (d)-(f) Corresponding $A_1$(LO) mode fitted with line shape analysis. Vertical dashed line is a guide to eye for the observed blue shift of $A_1$(LO) mode.

Figure 4. Variation of free carrier density and carrier mobility with respect to Ga/N ratio in GaN NWs.

Figure 5. Typical photoluminescence spectra of GaN NWs of samples (a) S1, (b) S2 and (c) S3, collected at room temperature and their corresponding low temperature (80K) spectra (d)-(f) respectively.

**Supplementary Information**

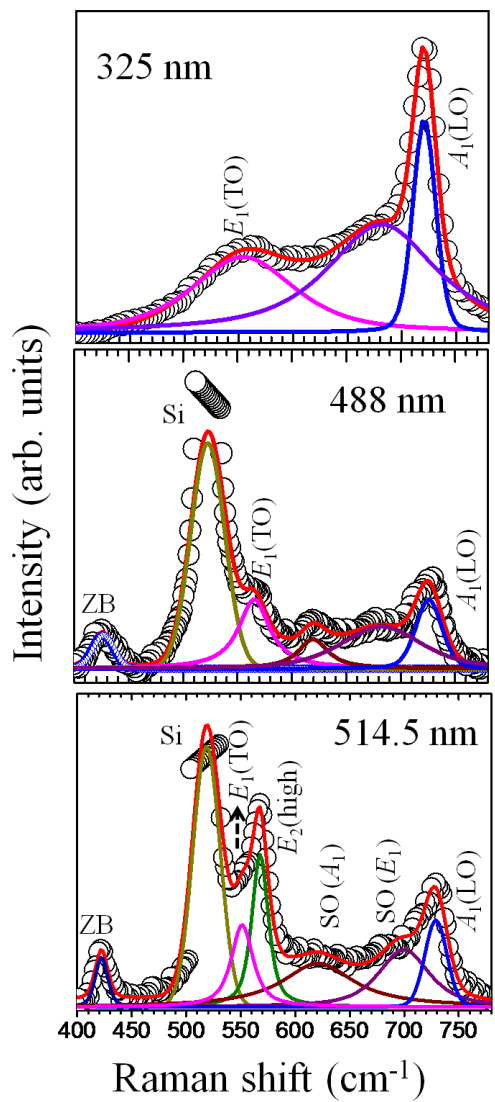

Figure S1. Comparison of typical excitation (514.5, 488, and 325 nm) dependent Raman spectra of nanowires of sample S3.

The excitation energy dependency of electron-phonon interactions in the GaN nanowires (NWs) of sample S3 was studied by collecting the Raman spectra of NWs using three excitation wavelengths 514.5, 488, and 325 nm for understanding the presence of Fröhlich interaction (FI) in the sample. A comparison of the typical Raman spectra of sample S3, excited by three wavelengths is shown. The relative intensity ratios of ($I_{E1-TO}/I_{A1-LO}$) are estimated to be 1.21, 1.00, and 0.35, respectively for the three wavelengths. The values clearly show that the longitudinal optical (LO) phonons are strongly interacted by the combination of FI and deformation-potential (DP) interactions when the excitation source is close to resonance condition of GaN.